\documentclass[pra,twocolumn,preprintnumbers,amsmath,amssymb,nofootinbib]{revtex4}
\usepackage{graphicx}
\usepackage[T1]{fontenc}
\usepackage[ansinew]{inputenc}
\usepackage{ae,aecompl}
\usepackage{enumerate,amsthm,amsmath,amssymb,color,float}
\newcommand{\ket}[1]{\vert#1\rangle}

\begin{document}
\title{Contexts, Systems and Modalities: a new ontology for quantum mechanics}
\author{Alexia Auff\`eves$^{(1)}$ and Philippe Grangier$^{(2)}$} 
\affiliation{
(1): Institut N\' eel$,\;$BP 166$,\;$25 rue des Martyrs$,\;$F38042 Grenoble Cedex 9$,\;$France \\
(2): Institut d'Optique, 2 avenue Augustin Fresnel, F91127 Palaiseau, France
 }
\begin{abstract}

In this article we present a possible way to make usual quantum mechanics fully compatible with physical realism, defined as the statement that the goal of physics is to study entities of the natural world, existing independently from any particular observer's perception, and obeying universal and intelligible rules. Rather than elaborating on the quantum formalism itself, we propose to modify the quantum ontology, by requiring that physical properties are attributed jointly to the system, and to the context in which it is embedded. In combination with a quantization principle, this non-classical definition of  physical reality sheds new light on counter-intuitive features of quantum mechanics such as the origin of probabilities, non-locality, and the quantum-classical boundary.

\end{abstract}
\maketitle

\section{Historical reminders}

It is well known that physicists, while they all agree about how to use Quantum Mechanics (QM), still all disagree about what it means, and even more about ``the real stuff" it describes: that is, its ontology.  However, let us take as a basis for an agreement  the two statements that (i) the quantum formalism is correct, and (ii) the quantum ontology must agree with physical realism, which states that the goal of physics is to study entities of the natural world, existing independently from any particular observer's perception, and obeying universal and intelligible rules\footnote{Some philosophical background is given in the Annex.}. Then the disagreement quoted above finds its roots in the difficulty to make (i) and (ii) compatible, as it has been much debated in the literature \cite{EPR,Bohr,Heisenberg,Bell-EPR,Bell,Bell-exp,Moon,Contextuality,pg2,Mermin,PBR}, giving rise to many different interpretations of QM \cite{Frank}. 

The purpose of this article is to propose and discuss a way to have the quantum formalism (i) and physical realism (ii) both correct  and compatible, so that many different views and practices about QM can stay essentially unchanged. As usual, there is a price to pay, but the currency will be ontological:  it will be a subtle but deep change in  what is meant by physical properties, which should not any more be considered as properties of the system itself, but jointly attributed to the system, and to the context in which it is embedded (precise definitions will be given below). We will show also that this ontological change has strong links with quantization as a basic physical phenomenon, and that this combination can explain why QM must be a probabilistic theory. 

Let us start with the simple question: what is a quantum state~? 
Since current answers do not meet a large agreement between physicists, it may be useful to remember that before QM, ``classical" physicists used to work with physical states, that pertained to physical systems. Physical systems are entities of the natural world that can be isolated well enough to study them, measure their properties, i.e. ``ask them questions" (what is your position, mass, velocity...). The set of answers resulting from a given set of questions defines a physical state, that is, the ID card of the physical system. Once this ID card is known, the behavior of the system becomes perfectly predictable from 
the equations of motion, and answers to new questions are predictable as well\footnote{This is clearly an idealized view of classical mechanics, ignoring e.g. chaotic systems, but what matters is that classical mechanics do behave generically in such a way.}. In the particular case where the measured properties are constant of the motion, the same questions will  invariably give the same answers, whatever their total number, or the ordering of their sequence. 
This is consistent with attributing a physical state  to the system itself,  so that the system ``is" in that state, even if nobody is there to look at it  \cite{Moon}. This property is often thought as the core of objectivity, and more generally, corresponds to ordinary ontology, i.e., our usual way of seeing (classical) reality.

With the rise of QM, the notion of physical state was seriously shaken. In particular, it appeared impossible to obtain as many certain answers as possible questions, i.e. to get the full ID card of the system. At early times of the quantum theory, this resulted in the famous Einstein-Bohr debate \cite{EPR,Bohr,Bell-EPR,Bell-exp}, with opposite claims that either the physical state exists like in classical physics, but then QM would be incomplete \cite{EPR}, or that QM would be complete, but then the full physical state does not exist, at least not in a classical sense  \cite{Bohr}. Then many different interpretations of QM have been proposed, and our purpose here is not to discuss all of them \cite{Frank},  but rather to examine another, maybe more philosophical,  perspective: assuming physical realism, can we spell out a  physical reality suitable for QM, and what are the practical consequences of such an ontological approach? 

In Sec.  II,  we will  present and discuss  the current - often implicit - quantum ontology. Then in Sec.  III and IV, we will introduce our approach, closely related to  ``contextual objectivity"  \cite{Contextuality,pg2}. Important consequences will be presented in Sec. V to VIII, and some related philosophical considerations will be discussed in the Annex. 

\section{Usual quantum ontology}

\begin{figure}[t]
\begin{center}
\includegraphics[width = 8cm]{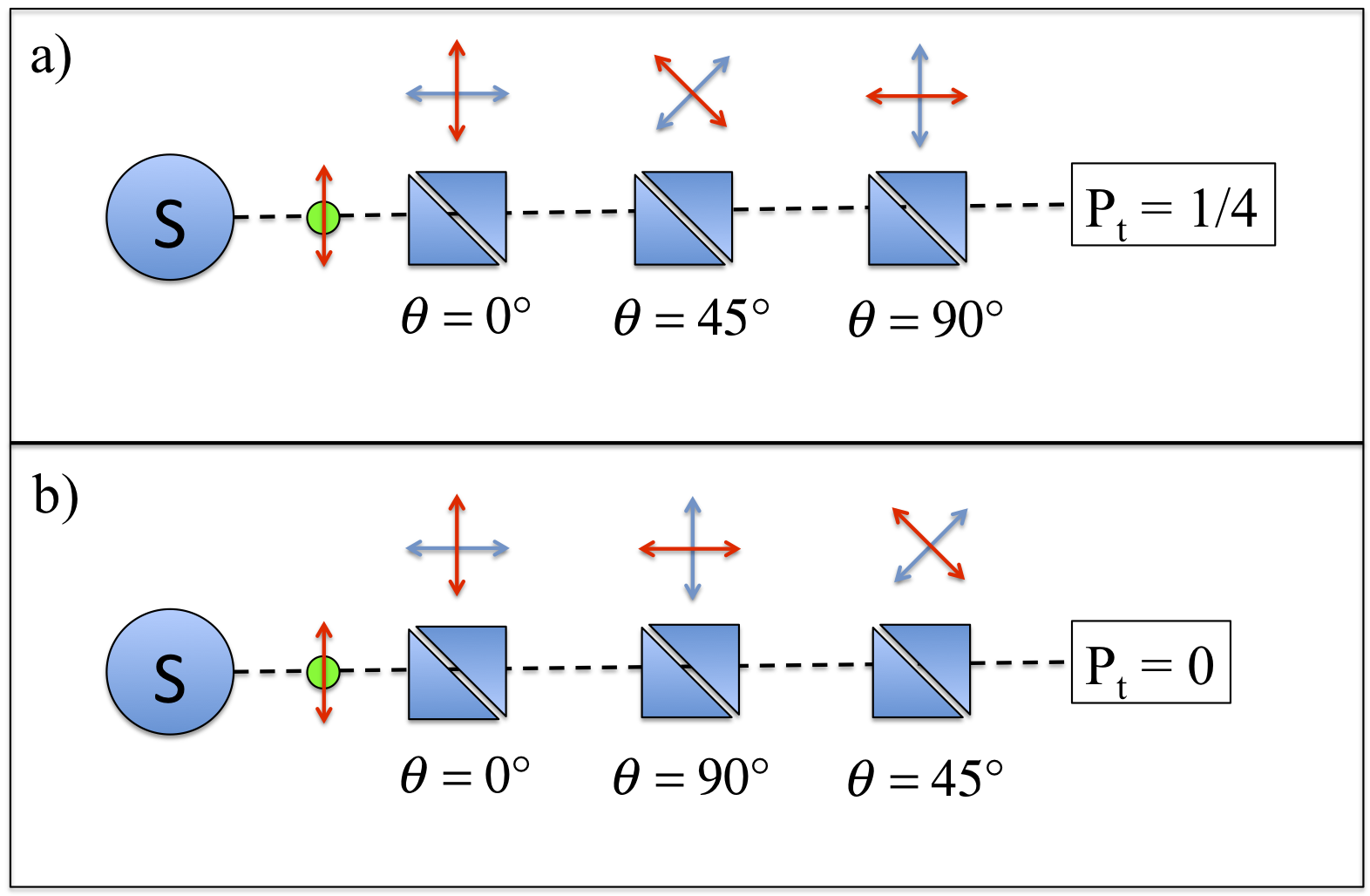}
\end{center}
\caption{
A H-polarized single photon is successively analyzed by three polarizers  oriented at either  0°, 45° and 90° (a) or  0°, 90° and 45°  (b). The overall probability of being transmitted is respectively either 1/4 (a) or 0 (b).}
\label{fig1}
\end{figure}

Nowadays in QM, one is accustomed to the idea that when studying a physical system, the full ID card, corresponding to the ensemble of answers to all the possible questions one can ask, is not accessible, and that the description must be restricted to the ``quantum state", quite different from what the ``classical state" used to be. A paradigmatic example of such quantum state is the photon polarization. It is usually measured with a polarizing beam-splitter, that can be rotated by a continuous angle $\theta$. By definition, the polarization is $\ket{H_\theta}$ (resp. $\ket{V_\theta}$) if the photon is transmitted (resp. reflected) with certainty by a $\theta$ oriented beam-splitter. In agreement 
with the Malus law  that applies for classical light, a photon of polarization $\ket{H_{\theta = 0}}$ sent in a $\theta \neq 0$ oriented beam-splitter will be either transmitted with probability $\cos^2(\theta)$, or reflected with probability $\sin^2(\theta)$. After the measurement, the photon is projected on the state $\ket{H_\theta}$ (resp. $\ket{V_\theta}$). 

This simple picture (Fig. \ref{fig1}) gathers important characteristics of quantum mechanics. First, there are indeed less certain answers than possible questions, meaning in that specific example that the photon polarization cannot be known with certainty for all $\theta$ angles. Second, results of successive measurements depend on the ordering of the sequence; for instance, it is not equivalent to measure the polarization successively in the basis $\{0°,45°, 90°\}$, or $\{0°,90°,45°\}$ - in other words, measurements do not commute\footnote{Classically one could do the same experiment with a polarized light wave, and also get different results according to the ordering of the polarizers. However the interpretation is completely different : the classical polarization ``has" an orientation in space, which can be determined by carrying out more measurements, whereas the photon can only give a probabilistic yes/no answer.}. 
These characteristics derive from the fact that 
measurement is a projection, from an initial state onto an eigenstate associated with a ``measurement context". 

Even if this conception is commonly spread, it holds many unsolved questions  that makes the understanding of quantum mechanics a vivid field of debates and interpretations, more than a hundred years after it was born \cite{Peres-Zurek,Bohm,GRW,Rovelli,Griffiths,RMP,c*,AZ,hardy,Zurek,Fine}. For instance, on the ontological side, what does the quantum state represent: our knowledge of reality, or reality itself - in other words, is the wave function epistemic, or ontic, or a bit of both \cite{PBR}~? What happens during a measurement, at the microscopic scale~? Where do probabilities come from, and how to justify them~? What happens during an EPR / Bell experiment carried out on entangled particles, and what is this ``spooky action at a distance" that affects Bob's photon, when Alice measures hers~? Last but not least, measurement is at the heart of the usual quantum description - but is it possible to describe the world out of any measurement process, and thus to  get rid of the so-called quantum-classical boundary \cite{Bell,Mermin}~?

In the following we propose an alternative ontology for quantum mechanics.  Let us emphasize  that an ontology, that aims to define ``what reality is", 
can never be {\it demonstrated}, but is rather a ``thinking framework",  which is chosen for both intuitive and practical 
reasons. Our strategy will therefore consist in going back to the roots of ontology, unveil and question our intuitions on what reality is, and eventually, take the freedom to choose a new direction.
In particular, in the same way that quantum physicists progressively gave up with the idea that the quantum state should consist of the full ID card of the system, here we will question the fact that the quantum state should pertain to the system alone. 
As detailed below, our perspective sheds new light on the set of conundrums reminded above, but it is unavoidable that other ontologies may also fit with the same mathematical theory - which one gives the ``best fit" is ultimately the choice to be made.

\section{System, context, and modalities}

To define an ontology, our main guideline is to start from the basic question: what can we be certain of~? More precisely, within the physical framework we are interested in, which phenomena can we predict with certainty, and obtain repeatedly~? 
Though a full  ``ID card'' in the classical sense  is  no more available as said above, certainty and repeatability of  
phenomena will allow us to identify some items of an ID card, by providing necessary conditions to be able to define a ``state''. Such an approach, supported by quantum experiments, has a clear relationship with the criteria for physical reality given by EPR \cite{EPR} -- but the following will be different. 

Our quantum ontology will thus involve three entities of different natures. First comes the \textbf{system}, that is - as stated above - a subpart of the world that is  isolated well enough to be studied. The system is in contact with other systems, that can be a measuring device, an environment - no need to be more specific at this point. The ensemble of these other systems will be called a \textbf{context}. A given context corresponds to a given set of questions, that can be asked together to the system. A set of answers that can be predicted with certainty and obtained repeatedly within such a context will be called a \textbf{modality}.  Given these definitions, let us bind them together by the following rule: {\bf In QM, modalities are attributed jointly to the system and the context. } This principle will be called ``CSM", referring to the combination of Context, System, and Modality\footnote{This is clearly an idealized view of how QM works, but besides agreeing with the usual formalism, there is abundant experimental evidence that individual quantum objects do behave this way.}.  As a set of certain and repeatable phenomena, a modality fulfills  the above conditions for the objective definition of a quantum state\footnote{With respect to the usual QM formalism, a modality corresponds to a pure state. We adopt here the usual view that statistical mixtures correspond to an extra layer of classical probabilities, added over a truly  quantum structure provided by pure states. We note also that the crucial idea of certainty and repeatability is associated with projective measurements. On the other hand, ``blurred" measurements (such as POVM's) may be very useful tools, but they don't provide fully predictable and repeatable results, and therefore they are not relevant for our purpose.}.
 
At this point we can emphasize that the context is classical, in the sense that no other context has to be specified to define its state. 
Note that here neither size, nor a quantitative criterium has been made to draw the quantum-classical boundary: 
the quantum vs classical behavior is only related to the CSM principle itself, i.e., to the very definition of a modality. 

Obviously, CSM applies to the photon example quoted above: the system is the photon, the $\theta$-oriented polarizer is the context, and the two possible exclusive answers/modalities in this context are either ``transmitted", or ``reflected". 
To predict with certainty if a photon will be transmitted or reflected, one has to know the modality, which includes the context it corresponds to - the angle of the polarizer in that case\footnote{In practice our definitions can be restricted to some sub-ensemble of contexts, relevant for some degrees of freedom of the complete physical system. What actually matters is that the transformations within the relevant set of contexts have the mathematical structure of a continuous group, see section IV.}.
In the CSM perspective, a photon does not ``own" a polarization, but the ensemble photon-polarizer does.  If the context is known, and if the system is available, a modality defined in this same context can be recovered without error.
This property has been exploited for years by quantum communication technologies, and provides the core of quantum cryptography protocols \cite{BB84}. Here, we have drawn the consequences of this behavior in ontological terms. 

The resulting ontology is clearly different from the classical one,  where it is expected that a state should ``exist" independently of any context to guarantee objectivity. But even if CSM is fundamentally non-classical, physical realism is not lost: it still pertains to the ensemble made of context, system, and modality. Objectivity, defined as the independence from any particular observer's perception, is still guaranteed, but {\it the ``object" is the system and the context, and its ``properties" are modalities \cite{Contextuality,pg2}}.

It might even be argued that the CSM principle should apply to classical ontology as well, since physical states always show up in a given context.  As mentioned in the introduction however, in classical physics the ordering of the questions, i.e. of the contexts that are successively in contact with the system, does not have any influence on the results, and all the questions can get a definite answer. As a consequence, the context can be forgotten, and the modality can be fully attributed to the system alone - giving birth to the ordinary classical ontology. On the contrary, in QM, the ordering in the succession of the questions has a strong influence on the answers obtained (see Fig. \ref{fig1}), and therefore  the context cannot be forgotten. Hence the non-commutation of measurements mentioned above is intimately related to the CSM principle; we will come back to this in the discussion below.

Finally, after quoting EPR at the beginning of this section, let us emphasize that the CSM principle is not foreign to  Bohr's view, as expressed in his answer to the EPR argument \cite{EPR,Bohr}: {\it ``The very conditions which define the possible types of predictions regarding the future behavior of the system constitute an inherent element of the description of any phenomenon to which the term {\bf physical reality}  can be properly attached". } 
In this sentence, Bohr explicitly states that despite being classical, the ``very conditions" (i.e., the context) must appear together with the system in the description of quantum phenomena. 
In the following, we show that the CSM principle  does not come  as a bolt from the blue,  but is actually tightly bound to a quantization principle.

\section{Quantization principle}

\begin{figure}[t]
\begin{center}
\includegraphics[width = 8cm]{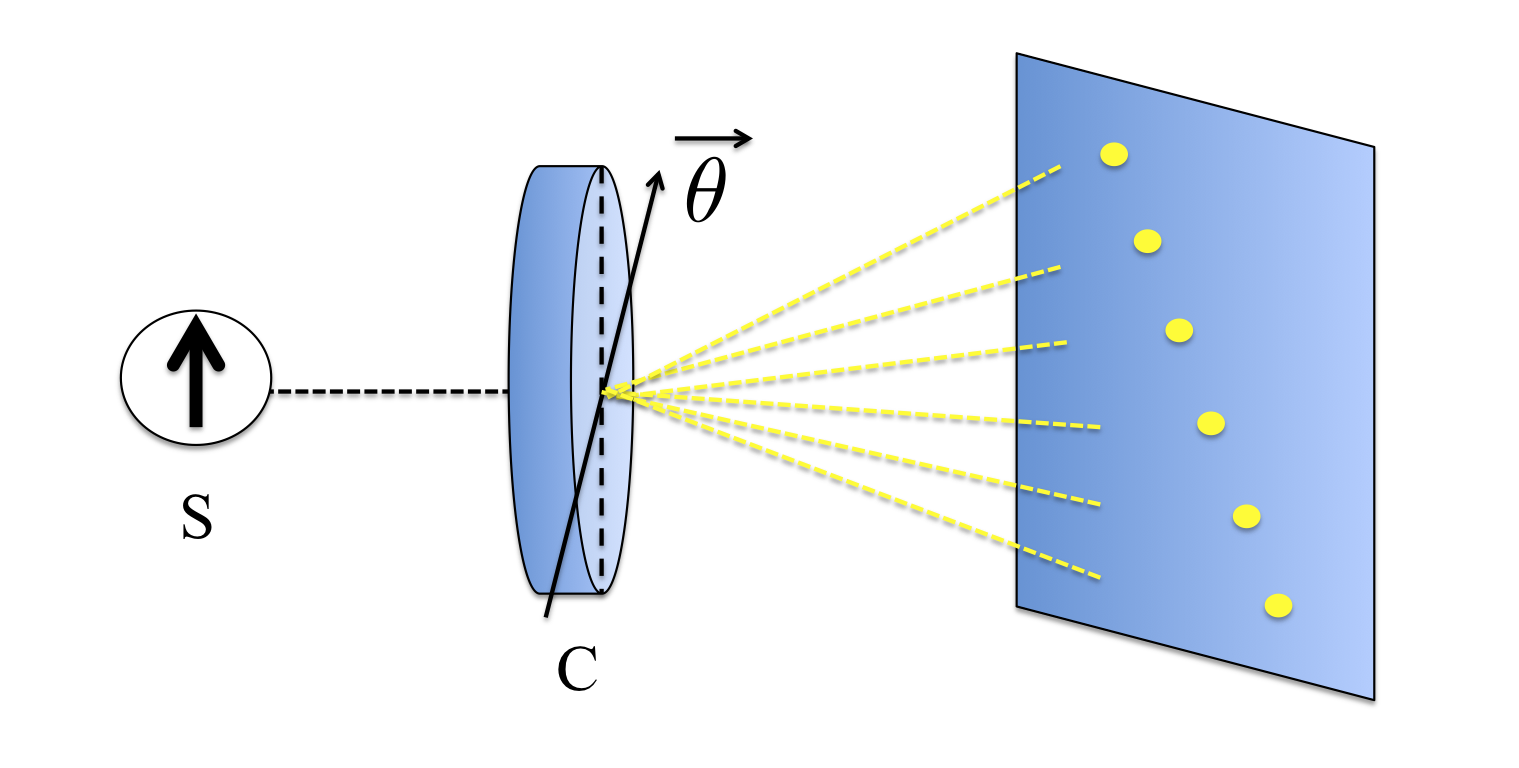}
\end{center}
\caption{
Illustration of the quantization principle for a spin 5/2  measured by a Stern-Gerlach apparatus. There is a continuous infinity of possible contexts, labelled by the angle $\vec{\theta}$ of the magnetic field gradient, but only N=6 mutually exclusive modalities. The number of spots, corresponding to the number of modalities for a given orientation $\vec{\theta}$, does not change while rotating the apparatus.
}
\label{Figure2}
\end{figure}

Let us consider the situation pictured in Fig.~\ref{Figure2}: a quantum system is described within a set of contexts, whose parameters can be changed continuously, whereas the measurement results (the modalities) are discrete and in a finite number\footnote{For continuous systems, see remark in section VII. The full consideration of infinite dimension deserves more discussion, and is postponed to another article \cite{new}.}. This is a quite general and canonical situation of quantum mechanics, which describes for instance spins interacting with a Stern-Gerlach apparatus, photons analyzed with polarizing beam-splitters, or quantum measurements on an ensemble of qubits. 

In a given context, the modalities are ``mutually exclusive'', meaning that if one modality is true, the others are wrong. On the other hand, modalities obtained in different contexts are generally not mutually exclusive: they are said to be ``incompatible", meaning that if one modality is true, one cannot tell whether the others are true or wrong.
This terminology applies to modalities, not to contexts, that are classically defined: changing the context results from changing the measurement apparatus at the macroscopic level, that is, ``turning the knobs''. 
These context transformations have the mathematical structure of a continuous group, that will be denoted $\cal G$: the combination of several transformations is associative and gives a new transformation, there is a neutral element (corresponding to no change of context), and each transformation has an inverse. In the following, we will thus consider infinitely many contexts, all related by the continuous group $\cal G$, and belonging to a set $\cal C$.

These definitions allow us to state the following quantization principle:

{\bf 
(i) For each well-defined system and context, there is a discrete number N of mutually exclusive modalities. This number N does not depend on any particular context within the set $\cal C$.

(ii) Modalities, when defined in different contexts within $\cal C$, are generally not mutually exclusive, and they are said to be ``incompatible''.}

Otherwise stated, whereas infinitely many questions can be asked, corresponding to all possible contexts, only a finite number N of mutually exclusive modalities can be obtained\footnote{This principle is reminiscent of other approaches which bound the information extractable from a quantum system  \cite{Rovelli,AZ}. However, in the realist perspective we chose, quantization has not a purely informational character, but characterizes reality itself.} (Fig.~\ref{Figure2}). 
An essential consequence is that it is impossible to get more details on a given system by combining several contexts, because this would create a new context with more than N mutually exclusive modalities, contradicting the above quantization principle. The resulting impossibility to define a unique context where all modalities are mutually exclusive  also makes that the succession of observed modalities depends on the sequence of contexts applied to the system (Fig.~\ref{fig1}). As mentioned above,  it is therefore forbidden to attribute definite physical properties to the system alone, in agreement with the CSM principle. Note that in our perspective, CSM does not show up because of some practical reasons pertaining to the measurement protocol, but is  intimately linked to the quantization principle. In the following, we revisit some of the main features  of quantum theory within this new framework.

\section{Probabilities}

As a first important conceptual consequence, we argue that quantum mechanics must be a probabilistic theory, not due to any ``hidden variables", but due to the ontology of the theory. The argument runs as follows: let us consider a single system, two different contexts $C_1$  and $C_2$, and the associated modalities $M(C_1,n)$ and $M(C_2,m)$, where $n$ and $m$ go from 1 to N. As said above,  the quantization principle forbids to gather all the modalities $M(C_1,n)$  and $M(C_2,m)$ in a single set of 2N mutually exclusive modalities. Looking for instance at photon polarization (N=2), it impossible to know  with certainty one the four possible results for the photon going through a polarizer oriented at $0^{\circ}$, {\bf and} through a polarizer oriented at $45^{\circ}$. Therefore the only relevant question to be answered by the theory is: if the initial modality is $M(C_1,n)$ in context $C_1$, what is the {\it conditional probability} for obtaining modality $M(C_2,m)$ when the context is changed from $C_1$ to $C_2$ ? Again, such a probabilistic description is the unavoidable consequence of the impossibility to define a unique context making all modalities mutually exclusive, as it would be done in classical physics. It appears therefore as a joint consequence of the quantization and CSM principles, i.e. that modalities are quantized, and require a context to be defined. 

Note that in the CSM picture as described so far, a ``measurement" is nothing but a change of context. As we will see  in section VII, this does not forbids to look for a more microscopic description of a measurement, e.g. by coupling the initial system to an ``ancilla". But this will simply lead to define some new and larger system and context,  in which the previous scheme will apply  again. 

\section{About the EPR argument}

Second, CSM sheds new light on the EPR argument \cite{EPR}. To show this, let us consider two spin 1/2 particles in the singlet state, shared between Alice and Bob. The singlet state is a modality among four mutually exclusive modalities defined in a context relevant for the two spins, where measurements of the total spin 
(and any component of this spin)  will certainly and repeatedly give a zero value. On the other hand, the singlet state 
is incompatible with any modality attributing definite values to the spin components of the separate particles in their own (spatially separated) contexts.

Now, let us assume that Alice performs a measurement on her particle, far from Bob's particle. Alice's result is random as expected, but what happens on Bob's side? Since Bob's particle is far away, the answer is simply that nothing happens. How to explain the strong correlation between measurements on the two particles? By the fact that after her measurement, Alice can predict with certainty the state of Bob's particle; however, this certainty applies jointly to the new context (owned by Alice) and to the new system (owned by Bob). The so-called ``quantum non-locality" arises from this separation, and the hidden variables from the impossible attempt to attribute properties to Bob's particle only, whereas properties must be attributed jointly to Alice's context and Bob's system. Getting them together is required for any further step, hence the irrelevance of any influence on Bob's system following Alice's measurement. Here the separation between context and system is particularly obvious and crucial, since they are in different places\footnote{In order to define a modality, i.e. 
a set of values that can be predicted with certainty, there is no need for a physical contact between the context and the system. Moreover, the context is not the whole physical environment, but only the classical data required to define the modality, e.g., the polarizer's orientation, which can be classically broadcasted. But checking the predicted results does require the actual measurement  to be done, i.e. the context and the system ``meeting" at the same place.}.

According to the above reasoning, after Alice's measurement on one particle from a pair of particles in a singlet state, the ``reality" is a modality for Bob's particle, within Alice's context.  But Bob may also do a measurement, independently from Alice, and then  the ``reality" will be a modality for Alice's particle, within Bob's context.   Does that mean that we have two ``contradictories" realities ?  Actually no, because these realities are contextual :  for instance Alice's modality tells that if Bob does a measurement in the same context as Alice, he will find with certainty a result opposite to Alice's one (given the initial singlet state). This statement is obviously true, as well as the one obtained by exchanging Alice and Bob. 
But if Bob does a measurement in another context (different from Alice's), then one gets  a probabilistic change of context for a N=2 system,  as described before.

If Alice and Bob both do measurements with different orientations of their analyzers, the simplest reasoning is to consider 
the complete context for both particles,  which is initially a joint context (with a modality which is the singlet state) and finally two separated contexts, 
again with 4 possible modalities due to the quantization postulate. Then this is now a probabilistic change of context for a N=4 system, 
again with the same result. 

It is interesting to write a few equations about these initial,  ``intermediate" and final  modalities,  
because this allows us to see more explicitly where CSM differs from Bell's hypothesis, even before the quantum formalism is introduced. 
So let us denote $a_i$, $b_j$  the modalities with results $i, \; j = \pm 1$ for some orientation (context)  $a$ for Alice, and $b$ for Bob. Given some ``hidden variables" $\lambda$, and using the vertical bar ``|"  as the usual notation for conditional probabilities $p(X|Y)$, Bell's hypothesis are :
$$p(a_i, b_j | \lambda) = p(a_i | \lambda) \; p(b_j | \lambda) $$
The equivalent CSM equations, given the initial joint modality $\mu$, are for Alice,  who knows $\mu$ and $a_i$
$$p(a_i, b_j | \mu) = p(a_i | \mu) \; p(b_j | \mu, a_i)$$
whereas they are for Bob, who knows $\mu$ and $b_j$
$$p(a_i, b_j | \mu) = p(a_i | \mu, b_j) \; p(b_j | \mu).$$
So these equations clearly differ from Bell's hypothesis, though there is no action at a distance, and no faster than light signalling. However, there is some non-locality, in the sense that the result on one side depends on the result on the other side; but this is only through a (local) redefinition of the context, not through any influence at a distance onto the remote particle. Again, it is essential here to consider that the modality belongs jointly to the particle(s) {\bf and} to the context, and not to the particle(s) only, otherwise one would be lead to Bell's hypothesis. 

Another important consequence is that  if Alice and Bob both do measurements, their realities must ultimately agree together, since there will be a unique final modality $(a_i, b_j)$. Therefore their predictions must also agree together, and one must have 
$$p(a_i, b_j | \mu) = p(a_i | \mu) \; p(b_j | \mu, a_i)  = p(a_i | \mu, b_j) \; p(b_j | \mu)   $$
These equations are just the same as the ones we would obtain by the usual ``instantaneous reduction of the wave packet", though in our reasoning there is no wave packet, and no reduction, but only a measurement performed by either Alice or Bob on the known initial modality $\mu$. Even more, if we admit that $(\mu, a_i)$ is a new modality for Bob, and $(\mu, b_j)$  is a new modality for Alice, then $p(b_j | \mu, a_i)$  or  $p(a_i | \mu, b_j)$  cannot be anything else than the one-particle conditional probabilities, i.e. Malus law for polarized photons (see section II). 

So we get a simple explanation about  the famous ``peaceful coexistence" between QM and relativity, i.e. why quantum correlations are non-local, but also ``no signalling" (they  don't allow one to transmit any faster than light signal): this is because a modality requires both a system and a context, and for instance $(\mu, a_i)$ defines a modality for Bob's particle, by using {\bf only} Alice's data. Such a situation, though strongly non-classical, does not conflict with physical realism  or causality: in the CSM perspective, quantum non-locality is a direct consequence of the 
bipartite nature of (quantum) reality.


\section{Evolution of modalities}

As said above, a modality is fully predictable and reproducible as long as the context is not changed, and the system does not evolve. More generally, the system will evolve, according to some physical laws, and it may also interact with other systems. As usual in QM, interactions can be treated  by combining systems, within a larger context suitable to define appropriate modalities. Modalities for the combined systems may be obtained by defining separate modalities for each sub-system, or by defining ``joint" modalities, which are predictable and reproducible only in the combined context. Such modalities correspond to entanglement, as discussed in the above section about the EPR argument. For two systems with respectively $N_1$  and $N_2$  mutually exclusive modalities, it is natural to admit (or postulate) that the combined system will admit $(N_1 \times N_2)$  mutually exclusive modalities. 

Considering again one system (possibly combined), the consequence of its evolution will be that predictability  and reproducibility will be lost in the initial context. However, one expects that  the result of the evolution will be a new modality, obtained from the initial one through a deterministic evolution. Therefore there will be a new context, in which the result of a measurement will again be fully predictable and reproducible. As it was assumed before, this new context will be related to the initial one by some element of the group of contexts transformation $\cal G$, and this element, as well as the modality itself, should  be determined by the equations of motion.  

At this point, it is essential to note that in an actual measurement, the system typically interacts with other systems (ancillas), and gets entangled with them within a larger context. By naive application of the usual quantum ontology, one does not know how to ``stop this process", and in principle the entanglement builds up until the observer himself is absorbed in the wave function.  Nothing like that occurs in the CSM ontology, because it is postulated from the beginning that a measurement carried out in a given context on a system with N exclusive modalities can only give one of these modalities, with some probability. Even entanglement, as long as it remains a meaningful concept, makes sense only with respect to an external context, always required for defining (combined) modalities and using the quantum formalism.  Let us point out also that measuring a modality actually means performing an  ``ideal QND measurement \cite{QND}", or a ``full projective Von Neumann\footnote{This calculation was initially done in Chapter VI of \cite{JvN}, which considers the ensemble made of  I (quantum system) + II (ancilla) + III (observer's device), and shows that for a properly designed system-ancilla interaction (in modern terms, it should be a QND interaction \cite{QND}), the same result is obtained by applying the measurement between I and II, or between II and III. Such an approach fully  agrees with CSM, and is very far from a ``many worlds" point of view \cite{RMP}.} measurement". In this limit, which is idealized but fully relevant for our purpose, a measurement is equivalent to a ``state preparation": this is consistent with the idea that a modality is defined by the certainty in the initial context, and not by the uncertainty in the (yet unknown) next contexts.  

Therefore there is no ``measurement problem" in CSM, but rather a crucial issue, which is to set up a mathematical formalism able to calculate probabilities connecting  different modalities in different contexts. The next question is then: which mathematical object should we associate with the physical concept of a modality ?  This will be quickly discussed in the next section, and more thoroughly in a forthcoming article \cite{new}.

\section{Born's formula}

We understand now that QM must be a probabilistic theory, 
and the theorems by Kochen - Specker and Bell \cite{Bell,Bell-exp} strongly suggest 
 to find  an alternative to classical probability theory, unless one accepts that it becomes contextual and non-local. 
 A new probability theory is therefore desirable, and our framework perfectly fits \cite{new} with the usual postulate: Let us associate to each modality a (rank-one) projector in an Hilbert space, so that any set of N mutually exclusive modalities $M(C_i,n)$ in a given context $C_i$ is associated to a set of N mutually orthogonal projectors $\Pi(C_i,n)$ summing up to identity. 

Then Gleason's theorem \cite{Gleason} states that the only possible way to write the conditional probability $P[ M(C_2,m) | M(C_1,n) ]$ is as $\text{Trace}[ \Pi(C_2,m) . \Pi(C_1,n) ]$: this is the usual Born formula. Within the framework of separable Hilbert spaces (or von Neumann algebra), it is well known that our initial restriction to a finite dimension N can actually be lifted, recovering the usual (Dirac - von Neumann) formulation of QM\footnote{Though contexts play a central role in our construction, let us emphasize that it  is ``noncontextual" in the sense associated with Gleason's theorem: this just means that the same modality can be found in different contexts. 
As an example, consider a system of two spin 1/2 particles $\vec S_1$  and  $\vec S_2$, and define $\vec S = \vec S_1  + \vec S_2$. Using standard notations
for coupled and uncoupled basis,  the $|m_1=1/2, m_2=1/2 \rangle$ modality in the context $\{S_{z1},S_{z2} \}$ is the same as the $|S=1, 
m_S=1 \rangle$ modality in the context $\{\vec S^2, S_{z} \}$, though other modalities in the same  two contexts are different. 
} \cite{new}.

Since we have now reached the starting point of QM textbooks \cite{cct},  it appears that 
the usual structure of QM is fully compatible with our approach;
in particular rewriting physical quantities  as operators and states
as rays is straightforward. With respect to standard textbook presentations, the main differences are: 
\begin{itemize}
\item  quantum probabilities do not happen ``by chance" and have nothing to do with ignorance, but they are a consequence of the quantization postulate;
\item quantum non-locality has nothing to do with an ``action at a distance", but appears because a modality belongs to both a system and a context;
\item  there  is no ``measurement postulate", since it is already included in the definition of modalities. 
\end{itemize}

More precisely, in our approach  there is no ``wave function" developing upwards and utimately  branching into a ``many-world" universe \cite{Frank}, but only (non-classical) probabilities   connecting modalities appearing in different contexts. These probabilities are calculated using the standard quantum formalism, and thus they  may involve interfering paths, as usual; however,  there is no ``wave", but only a ``wavelike behaviour" due to the quantum way to calculate probabilities, through projections in an Hilbert space. In this ``physically realist" point of view,  the mathematical state vector (or wave function) should be carefully distinguished from the modality, i.e. from the phenomenon defining  the ``object" (see Annex).

\section{Conclusion}

We shall conclude with a few remarks. 
First, contextual objectivity \cite{Contextuality,pg2} allows for an ontology to QM, this is the joint reality of the context, system, and modalities (CSM). This leads to reinterpret quantum nonlocality as the situation where the context and the system are separated in space: though the certainty (modality) is present, it cannot be ``verified" or ``actualized" as long as the context and the system are not put together again. Such a situation has no conflict with physical realism, but never happens in classical physics, where the physical properties are carried by the system alone. 

Second, let us note that for many physicists, putting the context in the very heart of the theory implied an unacceptable ``shifty split" \cite{Bell,Mermin} between the quantum world (attributed to the system) and the classical world (of the context). A lot of efforts have been made to get rid of it, especially to make the classical world emerge from the quantum world, by attempts to describe  contexts within the quantum formalism. Such attempts may exploit the fact that there is a considerable  flexibility for defining the boundaries of the system, especially when considering that (weak or strong) measurements can be done by entangling the initial system with a ``meter" (or ancilla) system, e.g. by doing Quantum Non Demolition (QND) measurements \cite{QND}.
But in our approach, extending such  measurements to include the context is self-contradictory: even by adding more and more ``meters",
the system can never grow up to the point of including the context. This is because without the context, modalities cannot be defined, leaving the system as a fuzzy object including everything, quite unsatisfactory from the perspective of physical realism. 

The quantum-classical boundary has therefore a fundamental character, and QM was born from it, both from a physical and from a philosophical point of view, as it is discussed in more details in the Annex below. 
In a nutshell, the CSM approach presented here, without restricting the generality nor the applicability of QM, acknowledges the fact that, as a scientific discipline, QM ``can explain anything, but not everything" \cite{Peres-Zurek}.

As a final remark, Bohr's arguments in \cite{Bohr} were quite right, but perhaps failed to answer a major question asked in essence by EPR in  \cite{EPR}: can a physical theory be ``complete" if it does not provide an ontology that should be clearly compatible with physical realism~? Unveiling such an ontology is what we propose to do here.


\section*{Annex: Some philosophical remarks on the nature of physical reality. }

It should be clear that in present approach, the weight of philosophy  is  larger than in many other interpretations of QM. Since we are still doing physics, and not philosophy, we present in this Annex some elementary philosophical considerations, spelling out the conceptual shifts introduced by the CSM ontology. 

From a philosophical point of view, let us first emphasize again that we adopt the point of view of physical realism, telling that the purpose of physics is to study entities of the natural world, existing independently from any particular observer's perception, and obeying universal and intelligible rules. Therefore philosophical issues like the separation between subject vs object  are  definitely out of our scope: we are interested in defining ``objects" consistent with QM, and we claim this can be done -- though not in a ``naive" (classical) sense. 

Here we want to discuss an objection which can be made to CSM, and could also be made to Bohr's answer to EPR: in ordering to keep the above arguments consistent, the context seems to acquire a special status, ``evading'' the quantum description of reality which is being built. This objection can be answered relatively easily, but this requires a distinction between two notions of reality, that occur frequently in the history of philosophy. 

The first kind of reality is the ``ultimate material reality'', constituted by all the objects in nature which are,  from a scientific point of view,  made of particles, waves and all their combinations, giving rise to macroscopic objects. There is no need to be very specific about what this reality is made of, but it must have a major property: it does exist - i.e. it is external to our thinking - even if we know very little about it. And in some sense, we cannot know much about it, because it is just too complicated. It also has to be a ``global'' reality, because no part of it should play a particular role. It could also be called ``absolute reality'',  here we will call it ``ultimate reality''. 

The second kind of reality is ``empirical reality'', this is the reality of phenomena that are amenable to scientific knowledge. Empirical reality has two main properties: it is real, it does exist independently of the ``observer'', because it is obviously included in the ultimate reality; and it can be known, which means that it can be perceived and apprehended by observers, as knowledge must (also obviously) pertain to perceiving and thinking agents. Scientific knowledge of empirical reality is thus a synthesis of facts - ``what is really going on'' (in some subset of the ultimate reality), - and concepts, elaborated through the observation and formalization of what is going on. It is precisely this synthesis that produces ``understanding'', what we sometimes call the ``aha!'' effect. 

As said before, this distinction between ultimate and empirical realities is very old, probably as old as philosophy, but physicists often ignore it, and think that physics can address directly the ultimate reality, by defining, attributing and measuring properties that belong unconditionally to ``real objects''. Unfortunately, this way of thinking leads to a dead end as far as QM is concerned. 

Actually, physics always deals with empirical reality, not with ultimate reality. Its duty is to describe phenomena with mathematical tools, which will allow one to predict the values of measurable physical quantities. These measurements and their mathematical formalization take place in a framework where phenomena can occur and eventually be described and measured. In practice, this framework is the classical macroscopic world, and though this appears only as a practical requirement, it can hardly be escaped, due to the very nature of empirical reality. 
Let us emphasize that this statement does not restrict physics only to ``what can be perceived''. All along its history, physics has elaborated concepts that take an ontological value, such as atoms, and it is perfectly entitled to do so, because empirical reality is grounded on ultimate reality. Atoms are a very good example of such a progress: in the 19th century, they were introduced as abstract hypothetical entities with a strong explanatory power, then they were identified, and now they can easily be ``seen'' and manipulated at the individual level. More generally, physical concepts such as photon, electron, charge, mass, energy, fields... are also  entities required for describing the empirical reality in a synthetic way, referring again to the above mentioned synthesis of ``what is going on'' - the facts out there - and of its observation and formalization through physical, conceptual and mathematical tools that belong to science.

\begin{figure}[t]
\begin{center}
\includegraphics[width = 8.5cm]{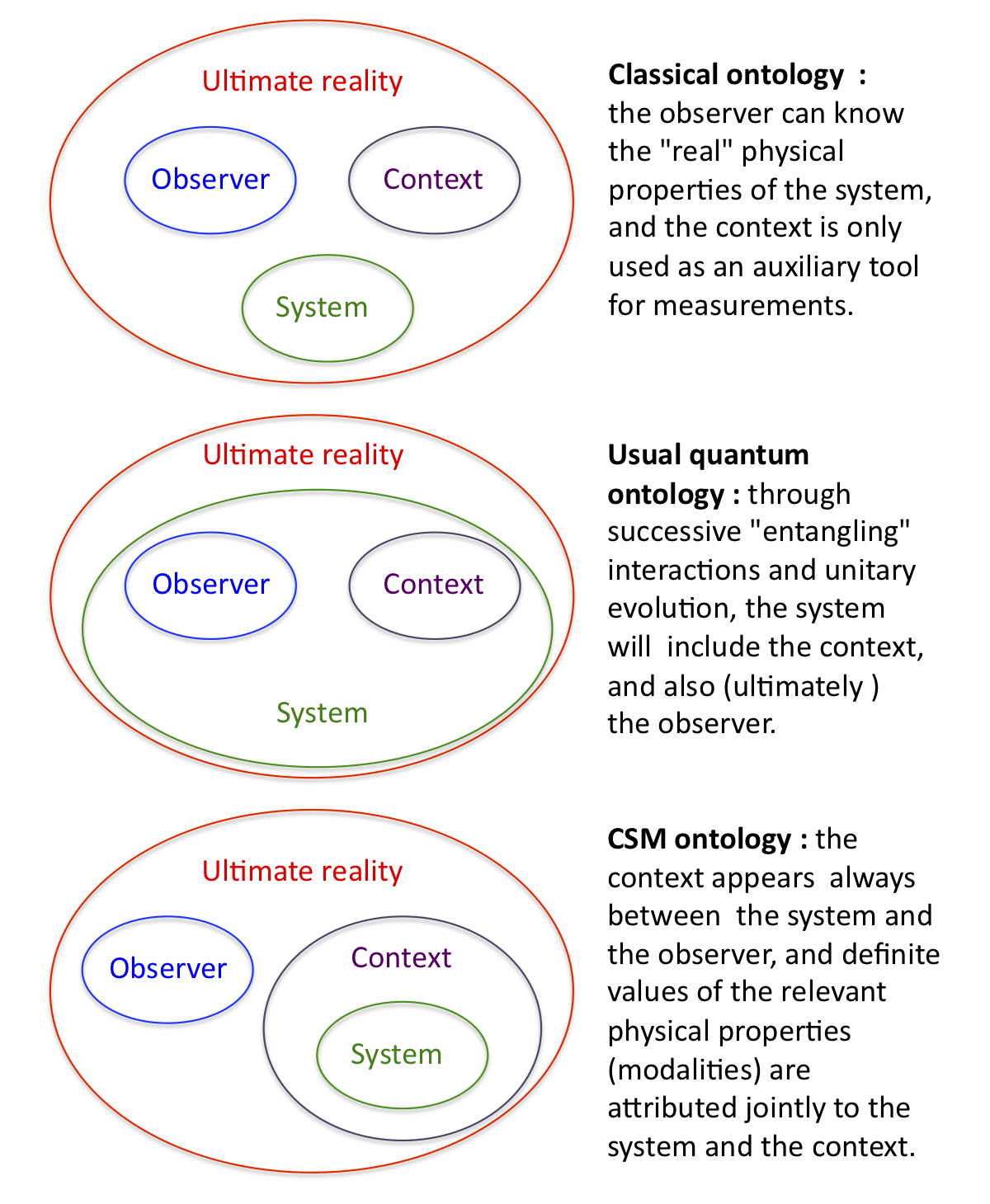}
\end{center}
\caption{Graphical representations of various ontologies discussed in the text. The CSM ontology is a much better
basis for physical realism than the usual quantum ontology. }
\label{fig2}
\end{figure}

Now we can consider again the difference between classical and quantum mechanics (see Fig. \ref{fig2}). In both, one deals with empirical reality grounded on ultimate reality, but in classical physics, one can easily get the delusion that knowing the state of the system is knowing directly the ultimate reality. In quantum mechanics, this is clearly wrong, because empirical reality must be mediated by a classical context, and the latter cannot be ignored. The context is always part of the ultimate reality, and its own very details in terms of particle, fields etc. certainly ``exist''. However, they are not relevant as far as the definition of a (CSM) modality is concerned: here the context is only considered as a necessary practical condition allowing the physicist to make measurements on  the system, which is a physically-grounded but abstract concept, like the atoms are. The context's role is to reveal a phenomenon, the modality, which must be accessible to the observer, in ordering for knowledge to take place. 

Given all that, a modality in CSM is essentially a ``phenomenon'' - a matter of fact - which involves a context (as a practical requirement) and a system (as a physically-grounded concept), and which provides measurement results, that can be known and reproduced with certainty. 
One should notice that most experiments do not give access to the ``full modality'', because of experimental imperfections: some properties may not be measured properly; experimental devices may add some noise, etc. However, the essential point is that, according to QM, modalities (i.e., pure quantum states) do exist as real phenomena, and the whole theory is based on that. In addition, the modality as a phenomenon is objective, i.e. it can occur anywhere, and requires no role for belief or for any agent's crucial presence. Though observation is part of scientific knowledge as a human endeavor, it is not required  for the phenomenal existence of a modality.

It should be clear also that the ``cut'' or ``split'' \cite{Mermin} is a requirement at the level of empirical reality, in ordering to specify the observed phenomenon, but at the level of ultimate reality, it does not imply that the macroscopic world is different in nature from the microscopic one. It rather means that in QM, macroscopic properties are required to describe phenomena, because the context cannot be ignored, due to the combination of the CSM and quantization postulates introduced above. Therefore for empirical consistency, the quantum system with its either mutually exclusive or incompatible modalities has to connect somewhere to the macroscopic world, where quantization does not show up at first sight. 

As a conclusion, a lot of trouble in QM results from a confusion between ultimate reality and empirical reality, associated with the classical delusion of ``speaking directly'' about ultimate reality. But this is no more possible in quantum mechanics, due to the empirical frontier imposed by the quantization postulate. Again, quantum mechanics can describe anything, but one should be very careful with attempts at using it to describe everything.

\subsection*{Acknowledgements}
The authors thank Nayla Farouki for essential contributions, especially in the Annex, and Francois Dubois, Franck Lalo\"e, 
Maxime Richard,  Augustin Baas, Cyril Branciard for many useful discussions. 


\end{document}